\documentclass[global,twocolumn]{svjour}
\usepackage{graphics}
\usepackage{epsfig}
\journalname{}

\def\jpb{J.\ Phys.\ B: At.\ Mol.\ Opt.\ Phys.\ }
\def\pra{Phys.\ Rev.\ A}
\def\pre{Phys.\ Rev.\ E}
\def\prl{Phys.\ Rev.\ Lett.\ }

\def\beq{\begin{equation}}
\def\eeq{\end{equation}}

\def\reff#1{(\ref{#1})}

\def\subsc#1{{\mbox{\rm\scriptsize #1}}}

\def\Wcmcm{\mbox{\rm Wcm$^{-2}$}}

\def\N3d{N_\subsc{3D}}

\def\vekt#1{\vec{#1}}

\def\vektr{\vekt{r}}

\def\vektR{\vekt{R}}

\def\vektE{\vekt{E}}

\def\vektp{\vekt{p}}
\def\vektP{\vekt{P}}

\def\rtilde{\tilde{r}}

\def\ptilde{\tilde{p}}

\def\hamop{{\cal{H}}}

\def\Edach{\hat{E}}

\def\energy{{\cal{E}}}

\def\pabl#1#2{\frac{\partial #1}{\partial #2}}

\def\calV{{{\cal V}}}

\begin{document}
\title{Small rare gas clusters in XUV laser pulses}
\author{D.~Bauer\thanks{E-mail: bauer@mbi-berlin.de}
}                     
%
%
\institute{Max-Born-Institut, Max-Born-Strasse 2a, 12489 Berlin, Germany}
\date{}
%
\maketitle

\begin{abstract}
Semi-classical molecular dynamics simulations of small rare gas clusters in short laser pulses of $100$\,nm wavelength were performed. For comparison, the cluster response to 800\,nm laser pulses was investigated as well. The inner ionization dynamics of the multi-electron atoms inside the cluster was treated explicitly. The simulation results underpin that at XUV wavelengths collisions play an important role in the energy absorption. The generation of the surprisingly high charge states of Xe atoms inside clusters, as they were observed in the free-electron laser experiment at DESY, Hamburg, Germany [Wabnitz {\em et al.}, Nature {\bf 420}, 482 (2002)], is due to the reduced ionization potential of atoms inside charged clusters, the ionization ignition mechanism, and collisions.
\end{abstract}

\section{Introduction} 
Clusters can absorb laser energy more efficiently than both gas targets and solids. This is because the cluster size is smaller than the skin depth so that all atoms experience the same laser field, despite the locally high particle density. Moreover, unlike solid targets there is no cold bulk that serves as a reservoir of cold electrons. Consequently, the laser-heated electrons cannot as easily escape from the cluster as they can from a solid surface. The electrons that absorb enough energy to leave the cluster as a whole contribute to the so-called ``outer ionization.'' Other electrons are removed from their parent atom or ion (``inner ionization'') but remain trapped by the positive background of the cluster. These electrons and the ions form a nanoplasma that hydrodynamically expands or Coulomb-explodes. In experiments at wavelengths $\geq 248$\,nm highly energetic electrons \cite{shao}, ions \cite{ditmirenature,ditmirePRA,springate}, photons \cite{mcpherson,ditmireJPB,teravet}, and neutrons originating from nuclear fusion \cite{zweibackDD} were observed (see Refs.\ \cite{review,reviewII} for reviews). 

Recently, Xe cluster experiments in the XUV regime have been performed at the DESY free-electron laser (FEL), Hamburg, Germany \cite{wabnitz}. Surprisingly, for laser intensities of a few times $10^{13}$, $100$\,fs pulse duration, and $98$\,nm wavelength multiply charged Xe ions were observed in the laser-cluster experiments even for small clusters whereas isolated Xe atoms loose only one electron under such conditions.  

The goal of this work is to shed some light on the mechanism underlying the generation of higher charge states inside clusters for laser parameters close to the FEL experiment. For that purpose molecular dynamics (MD) simulations have been performed. The inner-atomic dynamics is treated explicitly. The classical multi-electron atoms are rendered stable by introducing a momentum-dependent potential. Mechanisms contributing to inner ionization such as field ionization, collisional ionization, and even non-sequential ionization and ``shake-off'' processes are all self-consistently incorporated on a classical level. 

The paper is organized as follows. In Sec.~\ref{model} the numerical model is introduced. In Sec.~\ref{results} the numerical results are presented. Finally, we conclude in Sec.~\ref{concl}.

Atomic units (a.u.) are used throughout unless noted otherwise.

\section{MD model of rare gas clusters} \label{model}
The rare gas cluster consisting of $N_a$ atoms at the positions $\vektR_i$, $1\leq i \leq N_a$,  and $Z$ ``active'' electrons per atom at the positions $\vektr_j$, $1\leq j \leq ZN_a$ in a laser field $\vektE(t)$ is modeled by the Hamiltonian
\begin{eqnarray} \lefteqn{\hamop(\vektR,\vektP;\vektr,\vektp;t)=\sum_{i=1}^{N_a} \frac{\vektP_i^2}{2M} +  \sum_{j=1}^{Z N_a} \frac{\vektp_j^2}{2} } \label{hamiltonian} \\
&&  + \sum_{i=1}^{N_a}\sum_{j=1}^{Z N_a} \left(  V_H(\rtilde_{ij},\ptilde_{ij}) - \frac{Z}{\vert\vektR_i-\vektr_j\vert} \right)  \nonumber \\
&& + \sum_{i=1}^{N_a} \sum_{k=1}^{i-1} \left( V_{aa}(\vert\vektR_i-\vektR_k\vert) + \frac{Z^2}{\vert\vektR_i-\vektR_k\vert} \right)  \nonumber \\
&& + \sum_{j=1}^{ZN_a} \sum_{l=1}^{j-1} \frac{1}{\vert\vektr_j-\vektr_l\vert}  + \vektE(t)\cdot\left(\sum_{j=1}^{Z N_a} \vektr_j - Z \sum_{i=1}^{N_a} \vektR_i \right) . \nonumber
\end{eqnarray} 
Apart from the usual terms describing the kinetic energy, the Coulomb interactions, and the interaction with the laser field in dipole approximation, there are the additional potentials $V_{aa}$ and  $V_H$ accounting for the Lennard-Jones interaction between neutral atoms and the Heisenberg uncertainty principle for the electrons inside the atoms, respectively. The purpose of introducing these potentials will become clear when we now describe how the initial, unperturbed cluster configuration is built up.

(I) First, we seek an energetically optimal configuration for the neutral cluster without laser field. To that end we set $Z=0$ and
\beq V_{aa}(R_{ik})=D \left[ \left(\frac{a}{R_{ik}}\right)^{\!\!12} -  \zeta  \left(\frac{a}{R_{ik}}\right)^{\!\!6} \right] \eeq
where $\vektR_{ik}=\vektR_i-\vektR_k$, $R_{ik}=\vert\vektR_{ik}\vert$, and $\zeta$ is commonly set to either unity or two in the chemical physics literature. The potential minimum of $V_{aa}$ is at $R_{\min}=(2/\zeta)^{1/6}a$ and should be chosen close to the known nearest neighbor distance of the cluster under consideration.
By varying the parameter $D$ the overall strength of the Lennard-Jones potential $V_{aa}$ can be adjusted.

Starting from a random atom distribution with the nearest neighbor distances $ >  R_{\min}$ a local minimum in the energy landscape is obtained by propagating the atoms according
the equations of motion
\begin{eqnarray} \dot{\vektR}_i &=& \vektP_i/M, \\ 
\dot{\vektP}_i &=& - \pabl{\hamop}{\vektR_{i}}  - \nu\vektP_i 
\end{eqnarray}
with  a non-vanishing but small friction $\nu$ and the nuclear mass $M$ set to a small value so that relaxation occurs on an acceptable time scale.   For small clusters (say, $N_a < 10$) it is quite likely that also the {\em global} minimum is found in this way as long as one starts with a reasonable guess for the cluster structure.
Finding the global minimum for bigger clusters is far from trivial. 
Fortunately, the structures of Lennard-Jones clusters with $N_a < 150$ are available in the literature \cite{wales,waleswww}.

(II) Secondly, we seek an electronic configuration for the single atom with $Z$ active electrons and call this the ``mother configuration.'' Since ``classical atoms'' with more than one electron are generally unstable, a momentum-dependent potential
\beq
\calV(r,p,\xi,\alpha,\mu)=\frac{\xi^2}{4\alpha r^2 \mu} \exp\left\{ \alpha \left[ 1- \left(\frac{rp}{\xi}\right)^{\!\!4} \right]\right\}
\label{momdeppot} \eeq
is introduced \cite{kw} that enforces approximately the Heisenberg uncertainty relation when applied in the form
\beq 
V_H(\rtilde_{ij},\ptilde_{ij})=\calV(\rtilde_{ij},\ptilde_{ij},\xi_H,\alpha_H,\mu_{ei}) 
\eeq
where 
\beq \rtilde_{ij}= \vert \vektr_i-\vektR_j \vert, \qquad \ptilde_{ij}=\left|\frac{M \vektp_i- \vektP_j}{1+M}\right| . \eeq 
Here, $\mu_{ei}=M/(1+M)\approx 1$ is the reduced mass and $\rtilde_{ij}$ and $\ptilde_{ij}$ are the absolute values of relative distance and momentum, respectively.

The ``hardness parameter'' $\alpha_H$ governs how strictly the uncertainty relation $\rtilde_{ij}\ptilde_{ij}\geq\xi_H$ is fulfilled. Big values of $\alpha_H$ enforce it severely but also make the differential equations of motion stiff (which is numerically unfavorable). We have chosen $\alpha_H=2$. The parameter $\xi_H$ may be adjusted in such a way that the known essential ground state features of the system under study (e.g., the total energy or the ionization potentials) are properly mimicked. 

Although we neglect the spin in the current study it is worth noticing that the Pauli principle could be modeled in a similar way so that an atomic  shell structure is obtained \cite{cohen,cohen2000}. 

By solving the Hamilton equations of motion corresponding to \reff{hamiltonian} with a non-vanishing but small friction $\nu$, the ``mother configuration'' of atoms or molecules with only a few electrons can be easily found. For ``bigger'' atoms more advanced minimization routines have to be employed \cite{cohen,cohen2000}.  
Note that contrary to other classical trajectory Monte Carlo (CTMC) methods where the classical atom is modeled by an ensemble of electrons moving on Kepler orbits, in our case the groundstate is stationary, that is $\dot{\vektr}=\dot{\vektp}=0$ but, owing to the Heisenberg potential, $\vektp\neq 0$.

(III) Finally, the entire cluster is build up by taking $N_a$ randomly rotated electronic mother configurations and attaching them to the naked ions. The ion positions are known from the first step. Since now {\em all} electrons ``see'' each other (as well as all the other ions), another propagation with non-vanishing friction $\nu$ is required. During this step the electrons orient themselves in an energetically favorable way while the ions hardly move due to their huge mass $M$. Note that the cluster structure obtained in this way is not completely self-consistent because the ions are still sitting at the positions determined in step (I) where just the Lennard-Jones potential was effective. In a fully self-consistent procedure the semi-classical cluster should be assembled without a Lennard-Jones potential at all. However, this is numerically much more demanding and may lead to an ion distribution which is physically less reasonable than the Lennard-Jones structure after step (I).

We call the cluster configuration obtained after step (III) the ``cluster mother configuration.'' In order to obtain meaningful results concerning the ionization dynamics of the isolated atom or the cluster in a laser pulse, an ensemble of atoms and clusters had to be simulated. The members of such an ensemble were constructed by randomly rotating the cluster mother configuration. While this did not change the total energy of the cluster or single atom, of course, the orientation with respect to the laser polarization axis changed, and, hence,  so did the ionization dynamics. As expected, the ensemble-averaged entities converged more rapidly with respect to the time step for the integration of the equations of motion (a fourth order Runge-Kutta scheme was used) as the individual particle trajectories did. Convergence of the latter may be accelerated by reducing $\alpha_H$ or introducing a ``soft-core'' smoothing parameter for the electron-electron and electron-ion interaction, that is, e.g., replacing $\vert\vektr_j- \vektr_l\vert^{-1}$ by $[ (\vektr_j-\vektr_l)^2 + a_{ee} ]^{-1/2}$ since all this loosens the requirement of small time steps during close encounters of the particles.

\section{Results} \label{results}
We simulated small Xe clusters in intense laser fields. Each semi-classical Xe atom had $Z=3$ active electrons that were supposed to mimic three of the six 5p electrons of Xe. Treating more than three electrons per atom is not necessary for the modest laser intensities we consider in this work. 

The other simulation parameters were chosen $M= 131\times 1836$, $D=0.04$, $\zeta=1$, $R_{\min}=8.0$, $\alpha_H=2$, $\xi_H=1.75$. The ``mother configuration,'' i.e., a single Xe atom, consists of the three active electrons sitting on the corners of an equilateral triangle and the ion sitting in the center. The ion-electron distance was $1.6$ for all three electrons. The total energy of the configuration was $\energy_1=-2.30$ while the three ionization potentials were $I_1=0.32$, $I_2=0.80$, and $I_3=1.18$. The first three ionization potentials for the real Xe atom are $0.45$, $0.77$, and $1.18$. The parameter $\xi_H=1.75$ was chosen to optimize $I_2$ and $I_3$  since we are interested in {\em multiple} ionization and are particularly interested in the intensity regime where the removal of the first electron is given for granted for both the isolated atom and the atoms inside the cluster but higher charge states only occur in the cluster.
Spin was neglected for if we also introduced a Pauli-blocking potential $V_P$ of the form \reff{momdeppot} into \reff{hamiltonian} we would have modeled the Li atom (with its closed s-shell plus a loosely bound valence electron) rather than three of the six 5p electrons of Xe.

\subsection{Isolated atom} \label{ia}
\begin{figure}
\includegraphics[scale=0.42]{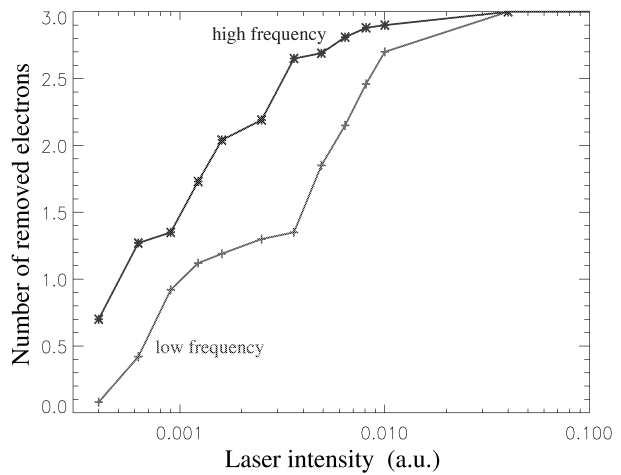}
\caption{\label{isolatom}  Charge state of the isolated model Xe atom after a $42$\,fs laser pulse of different peak intensity. Low frequency case: $\omega_l=0.057$, + symbols; high frequency case: $\omega_h=8\omega_l$, * symbols. Multiplication of the laser intensity by $3.5\times 10^{16}$ yields the laser intensity in the common units \Wcmcm.}
\end{figure}

In Fig.~\ref{isolatom} the charge state of the isolated Xe atom is shown vs the peak laser intensity of a $T=1764$ ($\approx 42$\,fs) laser pulse. The envelope of the electric field  was linearly ramped up (and down) over $T_r=331$ ($\approx 8$\,fs) while it was constant for the rest of the pulse. The two laser frequencies $\omega_l=0.057$ (corresponding to $800$\,nm) and $\omega_h=8\omega_l$ (corresponding to $100$\,nm) were used. 

As expected, for the same pulse duration and laser intensity the higher laser frequency ionizes the isolated atom more efficiently than the lower one.  For laser intensities $I<0.001$ ($\approx 3.5\times 10^{13}$\,\Wcmcm) it is unlikely to observe charge states higher than the singly ionized Xe. This is in agreement with the results of Wabnitz {\em et al.}\ \cite{wabnitz} where at an intensity $\approx 2\times 10^{13}$\,\Wcmcm\ for $12.7$-eV photons (corresponding to $8.2\,\omega_l$) and $\approx 100$\,fs pulse durations also higher charge states than Xe$^+$ were found to be absent for isolated atoms.  

\subsection{Small cluster}
\begin{figure}
\includegraphics[scale=0.42]{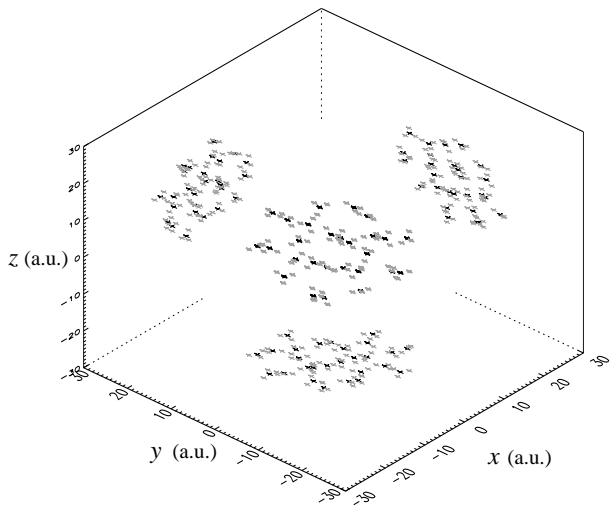}
\caption{\label{unpertcluster}  Unperturbed model Xe cluster configuration for $N_a=27$, $Z=3$, and $R_{\min}=8.0$. Ions are drawn in black, electrons in gray. Three projections are also shown. }
\end{figure}

In Fig.~\ref{unpertcluster} the unperturbed Xe model cluster consisting of $N_a=27$ triply charged ions (black) plus the $ZN_a=3\times 27$ electrons (gray) is shown. The ion structure is in agreement with Refs.\ \cite{wales,waleswww}. The electrons roughly maintain their triangular setup with respect to their parent ion. However, the triangles are in general not equilateral anymore due to the interaction with the neighboring atoms.  

The total energy of the unperturbed cluster amounts to $\energy_{27}=-63.3 < N_a \energy_1=-62.1$, i.e., the calculated cluster configuration is indeed energetically more favorable than $N_a$ isolated atoms. The binding energy per atom is $0.04$ which is a reasonable value for rare gas clusters.

\begin{figure}
\includegraphics[scale=0.65]{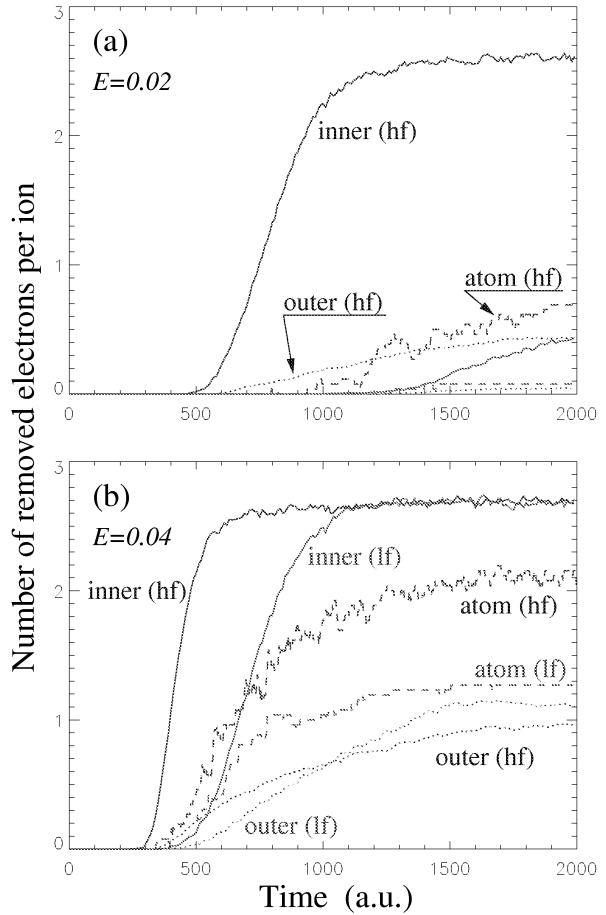}
\caption{\label{ionidegree}  Number of removed electrons per ion for the two laser intensities (a) $I_l=4\times 10^{-4}$ ($\Edach=0.02$) and (b) $I_h=1.6\times 10^{-3}$ ($\Edach=0.04$). The results for the isolated atom (dashed curves), the cluster outer ionization (dotted), and the cluster inner ionization (full) are plotted for the two frequencies $\omega_l=0.057$ (lf) and $\omega_h=8\omega_l$ (hf).  }
\end{figure}

From Fig.~\ref{isolatom} one can infer that for laser intensities $I$ less than $0.001$ no charge states $ > 1$ are expected in the case of isolated Xe atoms. Hence it is interesting to investigate whether higher charge states are observed for clusters. 
In Fig.~\ref{ionidegree} the results for the two laser frequencies $\omega_l=0.057$, $\omega_h=8\omega_l$ and the two laser intensities $I=4\times 10^{-4}$ ($\Edach=0.02$), $I=1.6\times 10^{-3}$ ($\Edach=0.04$) are collected for both the isolated atom and the cluster consisting of $N_a=27$ atoms. The laser pulse duration and shape was the same as in Sec.~\ref{ia}.  

The inner ionization of the cluster at the high frequency $\omega_h$ (indicated by ``inner (hf)'' in the plots) is relatively high for both laser intensities. The inner ionization was calculated by counting the number of electrons that are more than $8$\,a.u.\ away from their initial position and therefore must have left their parent ion. If recombination plays an important role inner ionization is overestimated by this method. However, by looking at the particle trajectories we inferred that it is unlikely that an electron which left its parent ion is {\em permanently} trapped by another ion. 

Inner ionization is rather low for $I_l$ and $\omega_l$. This is because an important prerequisite is missing for these parameters, namely the efficient removal of the outermost electron. It is clear that the very early inner ionization dynamics of clusters is the same as in the isolated atom case since there is not yet a strong influence of the {\em other} ions at that stage. In fact, the single atom ionization is small for $I_l$ and $\omega_l$ (see Fig.~\ref{ionidegree}a, dashed curve close to the bottom) so that the inner ionization of the cluster also remains quite modest.  This is different for the higher laser intensity $I_h$ (Fig.~\ref{ionidegree}b) where the isolated atom looses the outermost electron. Consequently, in the case of a cluster a space charge builds up that yields further inner ionization (``ionization ignition,'' \cite{rosepetru}).

Outer ionization (calculated by counting the number of electrons which are farther away than $r=30$) is always smaller than the corresponding ionization of the isolated atom for the parameters in Fig.~\ref{ionidegree} (see dotted curves). This means that a significant fraction of electrons that were removed from their parent ions do not make it to leave the cluster as a whole. This can be attributed to the low quiver energies of the electrons inside the cluster so that no collective electron dynamics sets in that could help to absorb additional laser energy and to overcome the strong backholding field of the cluster ions \cite{bauer}. 
Note that the ponderomotive energies for the parameters in Fig.~\ref{ionidegree} all lie between $4.8\times 10^{-4}$ and $0.12$ and are thus neither large compared to the photon energy nor large compared to the binding energies.

\begin{figure}
\includegraphics[scale=0.65]{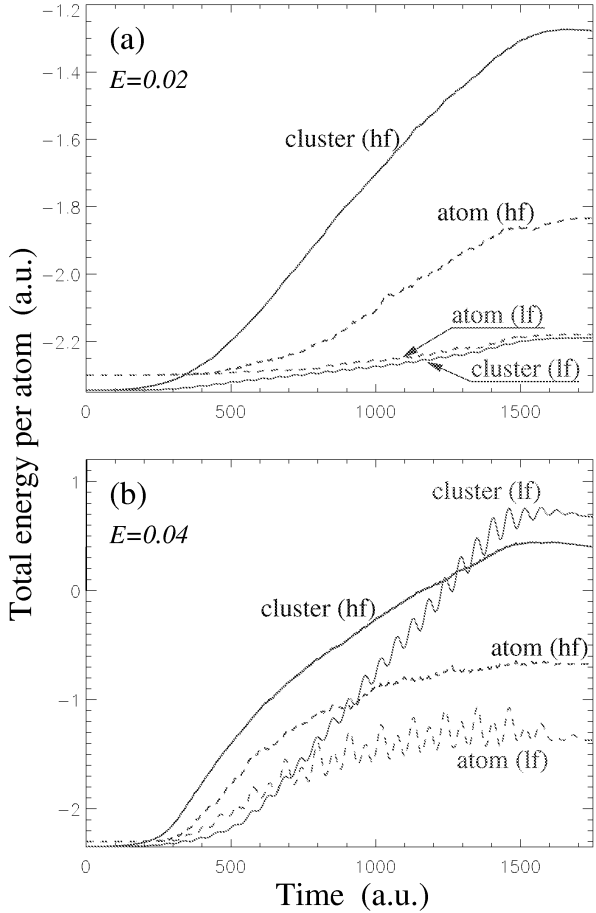}
\caption{\label{absorb}  Total energy vs time for the cluster (full curves)  and the isolated atom (broken curves). Laser and cluster parameters were the same as in Fig.~\ref{ionidegree}.}
\end{figure}

In Fig.~\ref{absorb} the total energy (per atom) $\energy(t)$ during the course of the laser pulse is plotted vs time. The laser and cluster parameters were the same as in  Fig.~\ref{ionidegree}. The net absorbed energy (per atom) $\vert\energy(T)-\energy(0)\vert$ is for both frequencies and laser intensities greater for the cluster than for the isolated atom. As already mentioned in the discussion of the ionization degrees shown in Fig.~\ref{ionidegree}, the absorbed energy is low for $I_l$ and $\omega_l$ owing to the rather small probability for single ionization.  For the higher intensity $I_h$ instead, slightly more energy is absorbed at the low frequency $\omega_l$. This matches with the increased outer ionization at low frequencies  (cf.\ Fig.~\ref{ionidegree}b). Moreover, the kinetic energy of the freed electrons depends on the laser frequency as well. 

\begin{figure}
\includegraphics[scale=0.33]{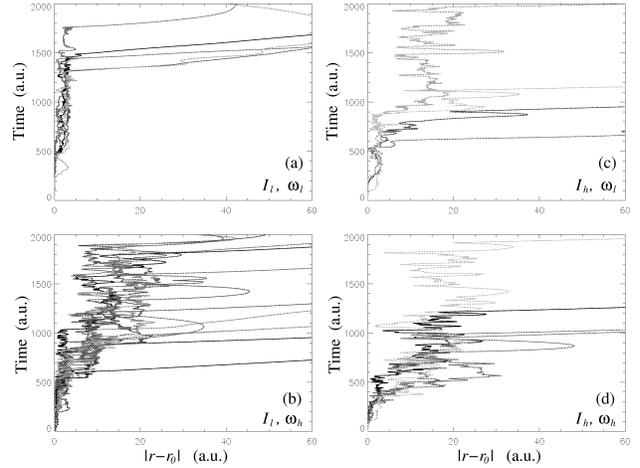}
\caption{\label{traject}  Sample trajectories of electrons which at the end of the laser pulse are more than $30$\,a.u.\ away from the cluster center. The absolute value of the distance to the electron's initial position, $\vert \vektr - \vektr_0\vert$, is plotted for the four different cases (a) $I_l,\omega_l$, (b) $I_l,\omega_h$, (c) $I_h,\omega_l$, and (d) $I_h,\omega_h$. }
\end{figure}

In order to understand the absorption mechanism at short wavelengths it is worth to look at sample electron trajectories. In Fig.~\ref{traject} one sees how electrons that at the end of the laser pulse are more than $30$\,a.u.\ away from the cluster center (and therefore contribute very likely to outer ionization) manage to escape from the cluster. As long as the distance from the initial position, 
$\vert \vektr - \vektr_0\vert$, is less than $\approx 8$\,a.u., the electrons can be considered bound with respect to their parent ion. Zig-zag motion between distances from $\approx 8$ to $\approx 23$ is due to collisions with {\em other} ions while electrons at distances $>  23$ left the entire cluster. 

For long wavelength and low intensity (a) one can see that the few electrons that made it to leave the cluster did so without collisions. This is because of the low probability for the removal of even the outermost electrons: most of the atoms remain neutral so that it is unlikely that the few freed electrons encounter an ion.

At higher frequency (b,d) it is seen from the erratic motion for $8< \vert \vektr - \vektr_0\vert < 23$ that the electrons experience several collisions with the cluster ions before they leave the cluster. Moreover, the electrons are temporarily trapped by other cluster ions.

In plot (c), for twice the laser intensity and the low frequency, collisions are observed as well. This is because both the quiver velocity and the thermal velocity are small so that electron-ion collisions are likely (see, e.g., \cite{mulser}). The outer electrons are removed efficiently at that laser intensity so that the ionization ignition mechanism starts (see the curve for inner ionization in Fig.~\ref{ionidegree}b). This indicates the eminent importance of the first freed electrons that trigger the entire cluster ionization process.

From the trajectories we deduce that the efficient absorption of laser energy, as it is seen in Fig.~\ref{absorb}, is due to collisional heating or, in other words, inverse bremsstrahlung. Similar conclusions were drawn recently by other authors using different methods \cite{preprsantra,preprrost}. The increased inner ionization is due to (i) the reduced ionization potential because of the neighboring ions, (ii) the field generated inside the cluster by {\em all} other ions, and (iii) collisions. All three mechanisms are included self-consistently in our numerical treatment which, unfortunately, implies that it is difficult to separate the different contributions and to identify their relative importance.

Although high charge states are generated inside the cluster during the course of the laser pulse it is not yet clear which charge states would arrive at the time-of-flight (TOF) detector in a real experiment. Upon the expansion of the cluster the electrons cool and some of them recombine, leading to a lower average charge state. Therefore we simulated the cluster disintegration as well. Since the cluster expansion happens on a picosecond time scale we reduced the ion mass to $M=50$ in order to make the simulation feasible \cite{remark}. While this mass reduction changes the time scale of the cluster expansion it should not affect too strongly the final energy and charge state distribution as long as $m/M\ll 1$.

\begin{figure}
\includegraphics[scale=0.38]{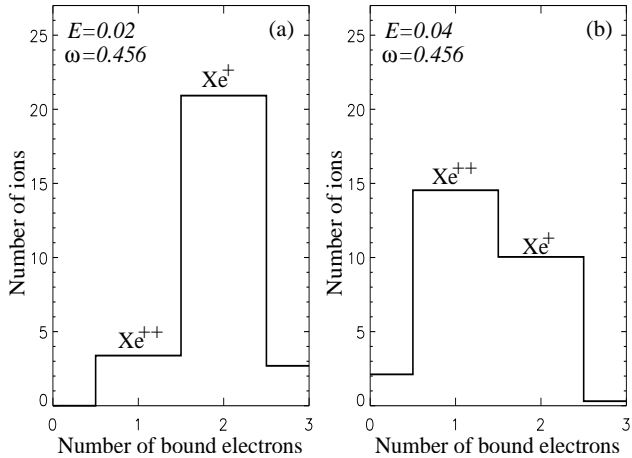}
\caption{\label{finalchargest} Charge state distributions after the expansion of the cluster for $\omega_h$ and (a) $I_l$, (b) $I_h$.}
\end{figure}

The result is shown in Fig.~\ref{finalchargest} for the high frequency $\omega_h$ and the two intensities $I_l$ and $I_h$. In Fig.~\ref{finalchargest}a it is seen that despite the plasma cooling a significant amount of Xe$^{2+}$ ions survive the cluster expansion. In the case of the higher intensity $I_h$ the doubly charged Xe ion is the most likely charge state and even triply ionized Xe atoms can be observed.

Comparison of Fig.~\ref{finalchargest}a with Fig.~1 in \cite{wabnitz} (TOF spectrum for $N=2$--$20$) suggests reasonable agreement. However, a precise comparison with the experiment \cite{wabnitz} is not straightforward because of the not exactly known experimental laser intensity, the varying sensitivity of the TOF spectrometer with respect to different charge states, and focus effects.

\section{Conclusions} \label{concl}
Molecular dynamics was used to investigate small rare gas clusters in short XUV and 800\,nm laser pulses. Inner ionization was treated explicitly. In order to render classical, stable multi-electron atoms possible, a momentum-dependent potential that accounts for the Heisenberg uncertainty principle was introduced. Higher charge states in clusters than in the isolated atoms  were observed, in accordance with experimental results. The mechanism underlying the efficient absorption of XUV laser energy was found to be inverse bremsstrahlung. 

The study of bigger clusters consisting of several hundreds or even thousands of atoms would be desirable. However, in order to keep the numerical simulations feasible one then has to give up the explicit treatment of the inneratomic dynamics.

\section*{Acknowledgments}
The author thanks A.\ Macchi for proofreading and valuable suggestions.   
This work was supported by the Deutsche Forschungsgemeinschaft and by the INFM Advanced Research Project CLUSTERS.
The permission to run our codes  on the Linux cluster at PC$^2$ in Paderborn, Germany, is gratefully acknowledged.


\begin{thebibliography}{}
\bibitem{shao} Y.\ L.\ Shao, T.\ Ditmire, J.\ W.\ G.\ Tisch, E.\ Springate, J.\ P.\ Marangos, and M.\ H.\ R.\ Hutchinson, \prl {\bf 77}, 3343 (1996).
\bibitem{ditmirenature} T.\ Ditmire, J.\ Tisch, E.\ Springate, M.\ Mason, N.\ Hay, R.\ Smith, J.\ Marangos, and M.\ Hutchinson, Nature {\bf 386}, 54 (1997).
\bibitem{ditmirePRA} T.\ Ditmire, T.\ Donnelly, A.\ M.\ Rubenchik, R.\ W.\ Falcone, and M.\ D.\ Perry, \pra {\bf 53}, 3379 (1996).
\bibitem{springate} E.\ Springate, N.\ Hay, J.\ W.\ G.\ Tisch, M.\ B.\ Mason, T.\ Ditmire, M.\ H.\ R.\ Hutchinson, and J.\ P.\ Marangos, \pra {\bf 61}, 063201 (2000).
\bibitem{mcpherson} A.\ McPherson, B.\ D.\ Thompson, A.\ B.\ Borisov, K.\ Boyer, and C.\ K.\ Rhodes, Nature {\bf 370}, 631 (1994).
\bibitem{ditmireJPB} T.\ Ditmire, P.\ K.\ Patel, R.\ A.\ Smith, J.\ S.\ Wark, S.\ J.\ Rose, D.\ Milathianaki, R.\ S.\ Marjoribanks, and M.\ H.\ R.\ Hutchinson, \jpb {\bf 31}, 2825 (1998).
\bibitem{teravet} S.\ Ter-Avetisyan, M.\ Schn\"urer, H.\ Stiel, U.\ Vogt, W.\  Radloff, W.\ Karpow, W.\ Sandner, and P.\ V.\ Nickles, Phys.\ Rev.\ E {\bf 64}, 036404 (2001).
\bibitem{zweibackDD} J.\ Zweiback, R.\ A.\ Smith, T.\ E.\ Cowan, G.\ Hays, K.\ B.\ Wharton, V.\ P.\ Yanovsky, and T.\ Ditmire, \prl {\bf 84}, 2634 (2000).
\bibitem{review} V.\ P.\ Krainov and M.\ B.\ Smirnov, Phys.\ Rep.\ {\bf 370}, 237 (2002).
\bibitem{reviewII} Jan Posthumus (ed.) {\em Molecules and Clusters in Intense Laser Fields} (Cambridge University Press, Cambridge, 2001). 
\bibitem{wabnitz} H.\ Wabnitz, L.\ Bittner, A.\ R.\ B.\ de Castro, R.\ D\"ohrmann, P.\ G\"urtler, T.\ Laarmann, W.\ Laasch, J.\ Schulz, A.\ Swiderski, K.\ van Haeften, T.\ M\"oller, B.\ Faatz, A.\ Fateev, J.\ Feldhaus, C.\ Gerth, U.\ Hahn, E.\ Saldin, E.\  Schneidmiller, K.\ Sytchev, K.\ Tiedke, R.\ Treusch, and M.\ Yurkov, Nature {\bf 420}, 482 (2002).
\bibitem{wales} David J.\ Wales, Jonathan P.\ K.\ Doye, J.\ Phys.\ Chem.\ A {\bf 101}, 5111 (1997).
\bibitem{waleswww} http://brian.ch.cam.ac.uk/
\bibitem{kw} C.\ L.\ Kirschbaum and L.\ Wilets, \pra {\bf 21}, 834 (1980).
\bibitem{cohen} James S.\ Cohen, \pra{\bf 51}, 266 (1995); \pra{\bf 57}, 4964 (1998).
\bibitem{cohen2000} James S.\ Cohen, \pra {\bf 62}, 022512 (2000).
\bibitem{rosepetru} C.\ Rose-Petruck, K.\ J.\ Schafer, K.\ R.\ Wilson, and C.\ P.\ J.\ Barty, \pra {\bf 55}, 1182 (1997).
\bibitem{bauer} D.\ Bauer and A.\ Macchi, \pra {\bf 68}, 033201 (2003).
\bibitem{mulser} P.\ Mulser, F.\ Cornolti, E.\ B\'esuelle, and R.\ Schneider, \pre {\bf 63}, 016406 (2000).
\bibitem{preprsantra} Robin Santra and Chris H.\ Greene, physics/0307058.
\bibitem{preprrost} Christian Siedschlag and Jan M.\ Rost, physics/0310123.
\bibitem{remark} Note that our numerical timesteps had to resolve the inneratomic dynamics.
\end{thebibliography}
\end{document}